\documentclass[useAMS,usenatbib]{mn2e}
\usepackage{graphics,epsfig,lscape,txfonts,color}
\usepackage[english]{babel}
\usepackage{longtable}

\newcommand{\mr}{\mathrm}

\newcommand{\hcm}[1]{$\times 10^{#1}$ cm$^{-2}$}

\def\ie{i.\,e.}                                      % i.e. (kursiv) \ie
\def\eg{e.\,g.}                                      % e.g. (kursiv) \eg

\def\gx339{GX\,339-4}
\def\h1743{H\,1743-322}
\def\xte{XTE~J1650-500}

\title[Relations between timing and spectral properties]{Relations between X-ray timing features and spectral parameters of Galactic black hole X-ray binaries}
\author[H. Stiele, T. M. Belloni, E. Kalemci, and  S. Motta]{H. Stiele$^{1}$\thanks{E-mail:
holger.stiele@brera.inaf.it}, T. M. Belloni$^{1}$, E. Kalemci$^{2}$, and S. Motta$^{3}$\\
$^{1}$INAF-Osservatorio Astronomico di Brera, Via E. Bianchi 46, I-23807 Merate (LC), Italy \\
$^{2}$Sabanc\i\ \"Universitesi, Orhanli-Tuzla, 34956 Istanbul, Turkey\\
$^{3}$European Space Astronomy Centre (ESAC)/ESA, PO Box 78, E-28691 Villanueva de la Ca\~nada, Madrid, Spain}
\begin{document}

%\date{Accepted 1988 December 15. Received 1988 December 14; in original form 1988 October 11}
\date{2012 December 3}

\pagerange{\pageref{firstpage}--\pageref{lastpage}} \pubyear{2012}

\maketitle

\label{firstpage}

\begin{abstract}
We present a study of correlations between spectral and timing parameters for a sample of black hole X-ray binary candidates. Data are taken from \gx339, \h1743, and \xte, as the \emph{Rossi X-ray Timing Explorer} (RXTE)  observed complete outbursts of these sources. In our study we investigate outbursts that happened before the end of 2009 to make use of the high-energy coverage of the HEXTE detector and select observations that show a certain type of quasi-periodic oscillations (type-C QPOs). The spectral parameters are derived using the empirical convolution model \texttt{simpl} to model the Comptonized component of the emission together with a disc blackbody for the emission of the accretion disc. Additional spectral features, namely a reflection component, a high-energy cut-off, and excess emission at 6.4 keV, are taken into account.  Our investigations confirm the known positive correlation between photon index and centroid frequency of the QPOs and reveal an anti-correlation between the fraction of up-scattered photons and the QPO frequency. We show that both correlations behave as expected in the ``sombrero" geometry. Furthermore, we find that during outburst decay the correlation between photon index and QPO frequency follow a general track, independent of individual outbursts. 
\end{abstract}

\begin{keywords}
X-rays: binaries -- X-rays: individual: GX 339-4, H 1743-322, XTE J1650-500 -- binaries: close -- black hole physics
\end{keywords}

\section{Introduction}
Most known black hole X-ray binaries (BHT) are transient. They are only observable during outbursts, as they are too faint to be detectable with present X-ray instruments during quiescence  \citep[see e.\,g.][]{1998ASPC..137..506G}. While a BHT is in outburst it evolves through different states, which show characteristic timing and spectral properties. The states can be identified with the help of the hardness intensity diagram \citep[HID;][]{2001ApJS..132..377H,2005A&A...440..207B,2005Ap&SS.300..107H,2006MNRAS.370..837G,2006csxs.book..157M,2009MNRAS.396.1370F,2011BASI...39..409B}, the hardness-rms diagram \citep[HRD;][]{2005A&A...440..207B}, and the rms-intensity diagram \citep[RID;][]{2011MNRAS.410..679M}. In general, outbursts begin and end in the low hard state (LHS) and there is a transition to the high soft state (HSS) in between. Three different types of low frequency quasi-periodic oscillations (LFQPOs) can be distinguished in BHTs \citep{2005ApJ...629..403C,1999ApJ...526L..33W}. In the LHS and the hard intermediate state \citep[HIMS; following the nomenclature of][]{2010LNP...794...53B} a specific timing feature named type-C QPOs can be observed \citep{2011BASI...39..409B}. 

Existence of a correlation between the QPO frequency and the photon index was first shown by \citet{1999ApJ...526L.101D}. They focused their discussion on the implications of the limited QPO frequency band on the limited change in the inner disc radius during transitions. \citet{2009ApJ...699..453S} studied correlations between the centroid frequency of QPOs and spectral parameters for a sample of eight BHTs observed with the Rossi X-ray Timing Explorer (RXTE). To obtain the spectral parameters RXTE/PCA (Proportional Counter Array) spectra in the 3.0 -- 50.0 keV range were fitted with the \texttt{bmc} model \citep{1997ApJ...487..834T}, which is hardwired to a Planck function. 

In this paper we investigate correlations between spectral parameters and timing properties for a sample of seven outbursts from three different sources. The spectra are fitted with a disk blackbody convolved with the \texttt{simpl} model \citep{2009PASP..121.1279S}, which is an empirical model of Comptonization. To model the broad spectral features due to reflection accurately, and to determine a high-energy cut off -- if present --, we included RXTE/HEXTE data in our study. We used only observations that showed type-C QPOs. A discussion of the correlation between photon index and centroid frequency for type-B QPOs observed in the 2010 outburst of \gx339\ can be found in \citet{2011MNRAS.418.1746S}. 

\begin{figure}
\resizebox{\hsize}{!}{\includegraphics[clip,angle=0]{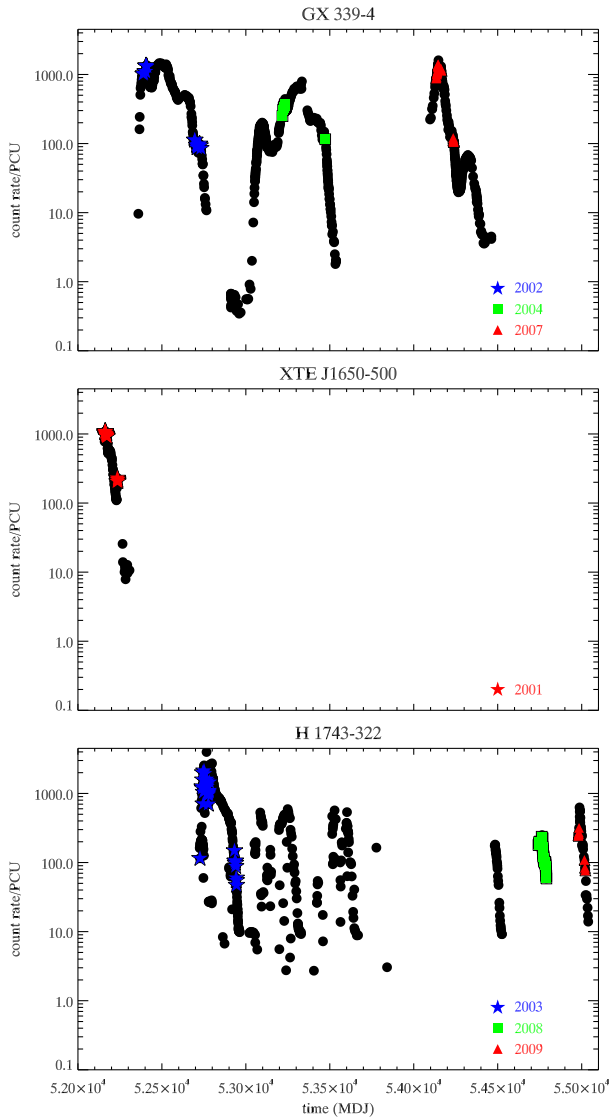}}
\caption{Long-term PCA light curves of \gx339\ (upper panel), \xte\ (middle panel), and \h1743\ (lower panel). Each point corresponds to an entire observation. The observations used in our study are marked with different symbols.}
\label{Fig:ltlc}
\end{figure}

\section[]{Observations and data analysis}
\label{Sec:obs}
In this paper we analysed archival RXTE observations of several outbursts of the black hole X-ray binary candidates \gx339, \h1743, and \xte, namely:
\begin{itemize} 
\item the 2002/03, 2004, and 2007 outbursts of \gx339
\item the 2003, 2008, and 2009 outbursts of \h1743
\item and the 2001 outburst of \xte, which is the only observed outburst of this source.
\end{itemize}

We selected these sources as RXTE observed complete outbursts of them, \ie\ neither the initial rise nor the decay to quiescence is missing; a substantial number of observations with type-C QPOs have been detected; and the HIDs were nicely q-shaped. In addition, RXTE observed several outbursts of \gx339 and \h1743, including outbursts at different luminosities and one outburst of \h1743 that did not go all the way to the soft state \citep{2010MNRAS.408.1796M,2011MNRAS.418.2292M}. Long-term PCA light curves are given in Fig.\ \ref{Fig:ltlc}.  
Outbursts that happened either partially or totally after the end of 2009 have been excluded from our study to avoid the introduction of systematic uncertainties in the spectral parameters due to the strong residuals in the HEXTE spectra that are related to the difficulties in determining the background contribution since technical problems occurred in the HEXTE detector \citep[see also][]{2012MNRAS.422..679S}.  
We investigated the timing and spectral properties, as described below.

\subsection{Timing analysis}
The values of the centroid frequencies and the classification of the QPOs for all three outbursts of \gx339 have been published in \citet{2011MNRAS.418.2292M}, and for the 2008 and 2009 outbursts of \h1743 in \citet{2010MNRAS.408.1796M}. The data of the 2001 outburst of \xte\ were analysed using the methods and classification criteria described in these papers. We used data from the Proportional Counter Array (PCA) to compute power density spectra (PDS) for each observation following the procedure outlined in \citet{2006MNRAS.367.1113B}. PDS production has been limited to the PCA channel band 0 -- 35 (2 -- 15 keV) and used 16 second long stretches of Event mode data.  
For observations which showed a quasi-periodic oscillation (QPO) we subtracted the contribution due to Poissonian noise \citep{1995ApJ...449..930Z}, normalised the PDS according to \citet{1983ApJ...272..256L} and converted it to square fractional rms \citep{1990A&A...227L..33B}. We determined the centroid QPO frequency, by fitting the noise components as well as the QPO feature with Lorentzians, following \citet{2002ApJ...572..392B}. PDS fitting was carried out within the standard \textsc{xspec} fitting package \citep{1996ASPC..101...17A} by using a one-to-one energy--frequency conversion and a unit response. The observations of \xte\ during outburst decay have been already analysed in \citet{2003ApJ...586..419K}. They used a 2 -- 30 keV energy range with 128 second long stretches\footnote{Using 16 second long or 128 second long stretches does not lead to any change in the derived QPO centroid frequency, as type-C QPOs are strong, narrow features in the PDS.}. For the 2003 outburst of \h1743 we took the QPO frequencies listed in Table~2A of \citet{2009ApJ...698.1398M}. They used the full bandwidth of the PCA instrument (2 -- 40 keV) and searched for QPOs in the 4 mHz to 4 kHz range with a sliding frequency window technique \citep{2002ApJ...580.1030R,2002ApJ...564..962R}.  
In the following, we focus on observations which QPOs we have classified as type-C. 
   
\begin{figure*}
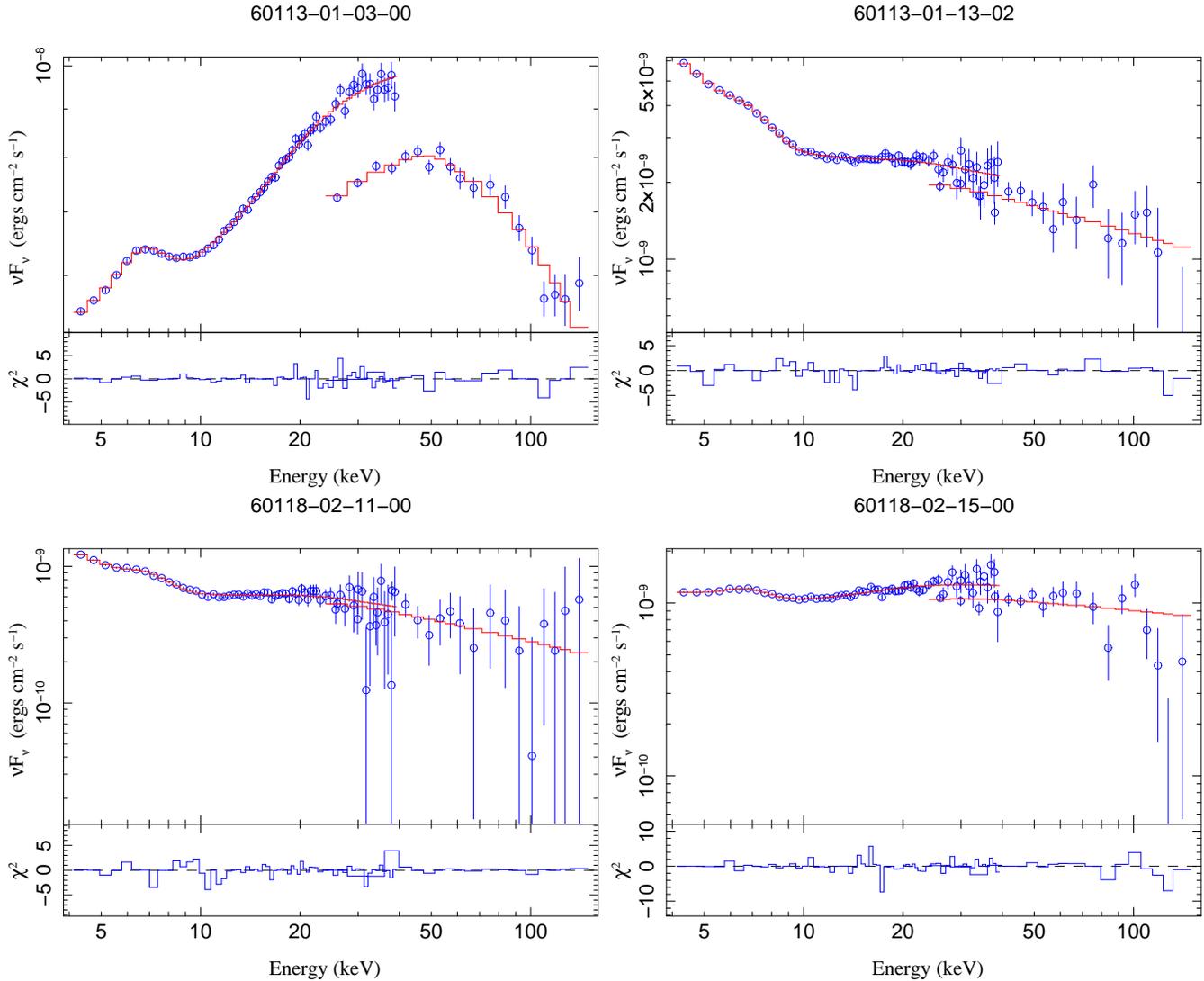

\resizebox{\hsize}{!}{\includegraphics[clip,angle=0]{spplot60113-01-03-00n_uf.ps}\hskip0.2cm\includegraphics[clip]{spplot60113-01-13-02n_uf.ps}}
\resizebox{\hsize}{!}{\includegraphics[clip,angle=0]{spplot60118-02-11-00n_uf.ps}\hskip0.2cm\includegraphics[clip]{spplot60118-02-15-00n_uf.ps}}
\caption{Unfolded spectra of \xte\ at the begin (upper left) and end (upper right) of the rise and at the begin (lower  left) and end (lower right) of the decay of its outburst.}
\label{Fig:spec}
\end{figure*}

\begin{figure}
\resizebox{\hsize}{!}{\includegraphics[clip,angle=0]{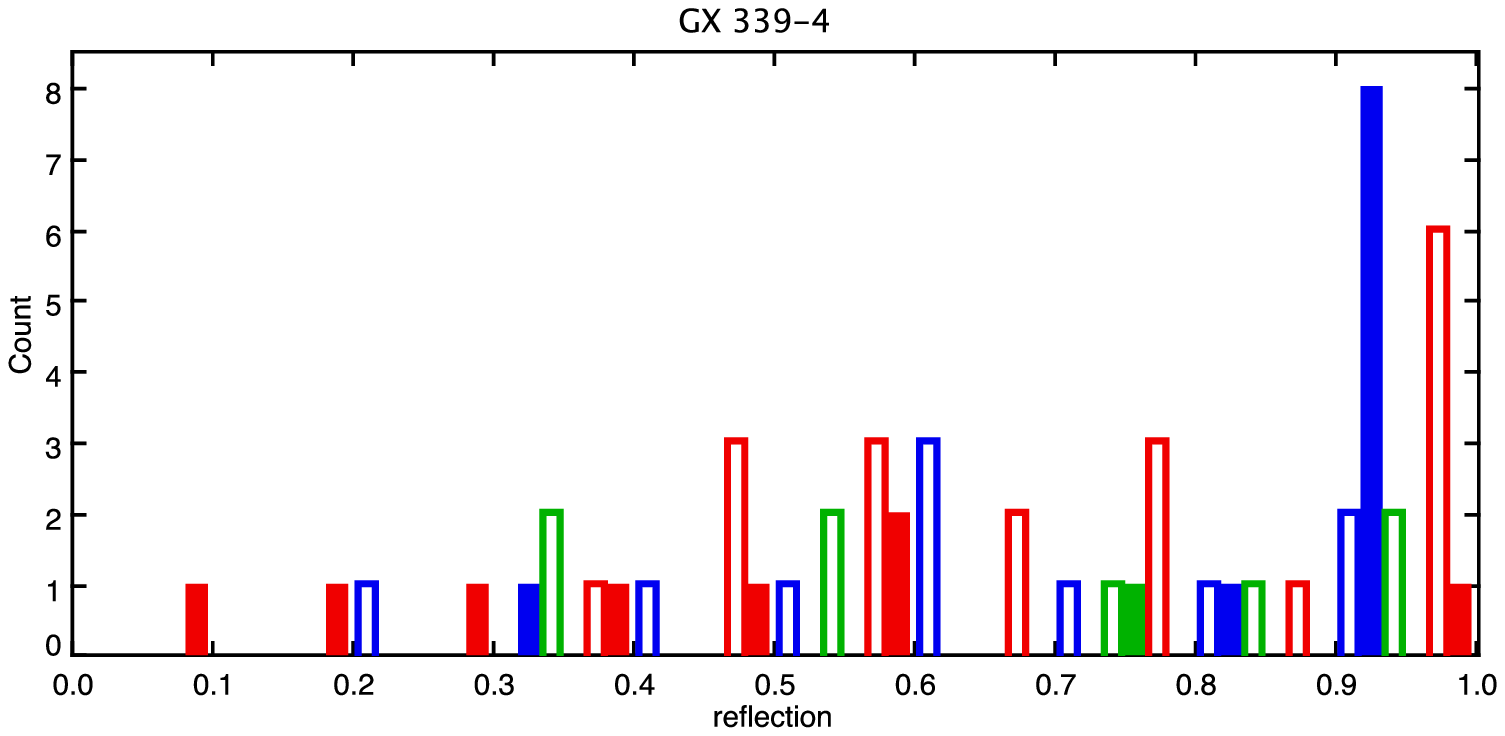}}
\resizebox{\hsize}{!}{\includegraphics[clip,angle=0]{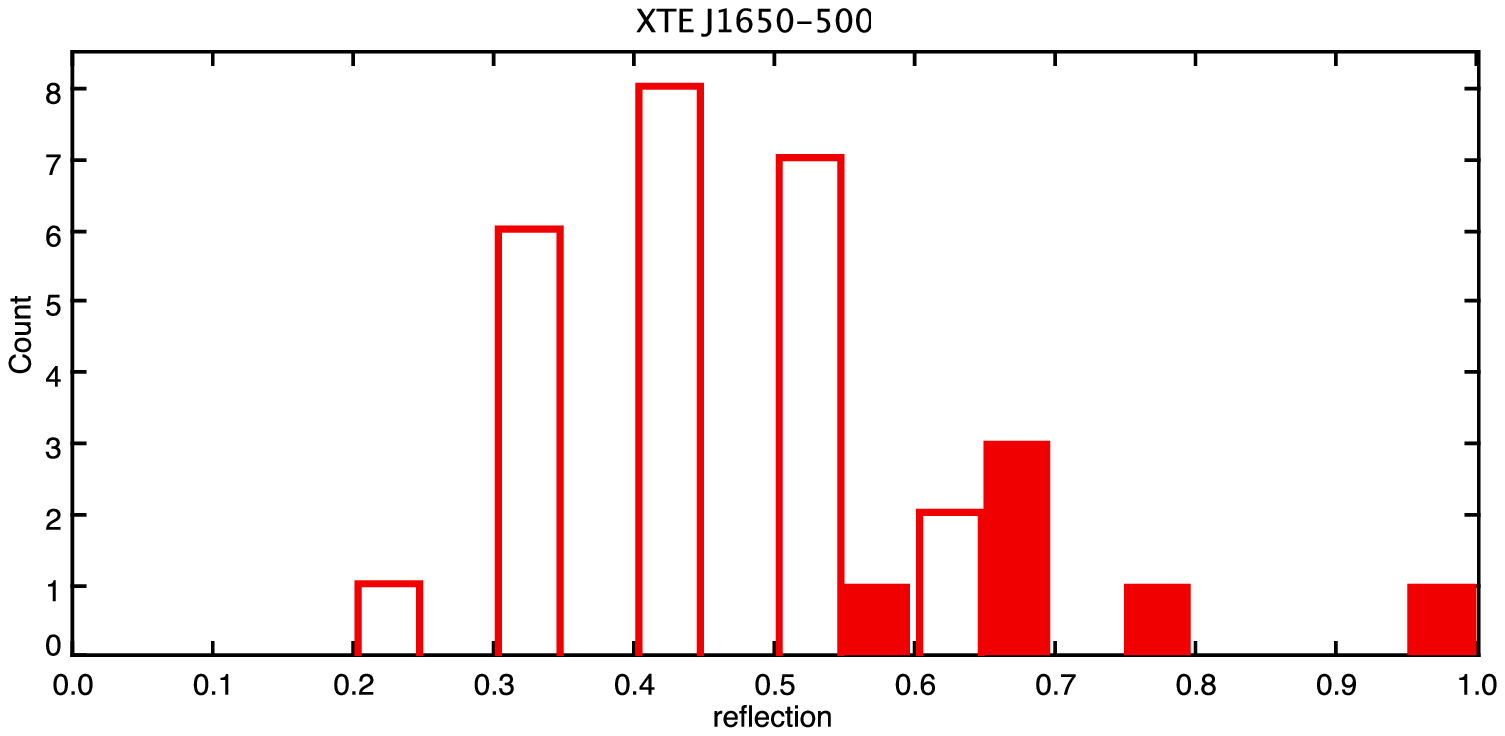}}
\resizebox{\hsize}{!}{\includegraphics[clip,angle=0]{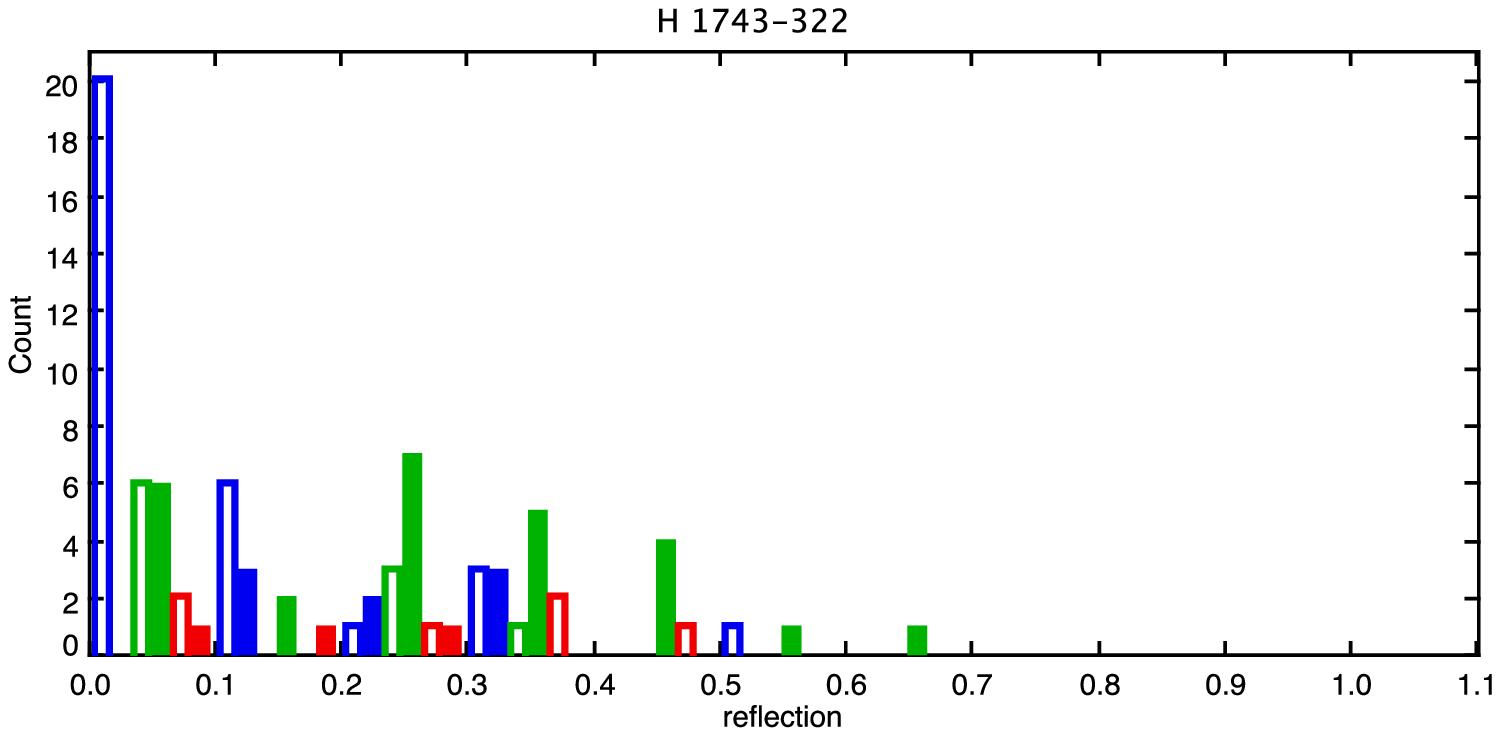}}
\caption{Distribution of the reflection values for \gx339\ (upper panel), \xte\ (middle panel), and \h1743\ (lower panel) with a binning of 0.1.\@ The mean error is about 0.15 (\gx339, \xte) to 0.2 (\h1743). Open bars denote observations belonging to the rise branch of an outburst, while filled bars mark observations of the decay branch. Different outbursts are marked with different colours (grey shades), using the same colouring scheme as in Figs.\ \ref{Fig:ltlc} and \ref{Fig:corr}.}
\label{Fig:refl}
\end{figure}

\subsection{Spectral analysis} 
For all observations with type-C QPOs we used the PCA Standard 2 mode (STD2), which covers the 2--60 keV energy range with 129 channels, and the HEXTE Standard mode, which covers the 15--250 keV energy range with 129 channels, for the spectral analysis. The standard RXTE software within \textsc{heasoft} V.~6.9 was used to extract background and dead-time corrected energy spectra for each observation. Solely Proportional Counter Unit 2 from the PCA was used since only this unit was on during all the observations. To account for residual uncertainties in the instrument calibration a systematic error of 0.6 per cent was added to the PCA spectra\footnote{A detailed discussion on PCA calibration issues can be found at: http://www.universe.nasa.gov/xrays/programs/rxte/pca/doc/rmf/pcarmf-11.7/}. For observations taken between 2001 and 2004 we used HEXTE data from cluster A, while for observations taken in 2007, 2008, and 2009 cluster B data have been used. 

Combined PCA+HEXTE spectra were fitted within \textsc{isis} V.~1.6.1 \citep{2000ASPC..216..591H} in the 4 -- 40 keV and 22 -- 200 keV range. We uniformly fitted the spectra with a partially Comptonized multi-color disc blackbody model, including foreground absorption. Four $\nu$F$_{\nu}$ spectra of \xte\ taken form the beginning and end of the rise and decay branch, respectively, are shown as exemples in Fig.\ \ref{Fig:spec}. \textsc{isis} ``unfolded spectra'' are independent of the assumed spectral model (\ie, the unfolding is done solely with the response matrix and effective area files; see \citealt{2005ApJ...626.1006N} for details). The plotted residuals, however, are those obtained from a proper forward-folded fit. The disc emission was approximated by the \texttt{diskbb} model \citep{1984PASJ...36..741M} and the \texttt{simpl} model \citep{2009PASP..121.1279S} was used for Compton scattering. The latter one being an empirical convolution model that converts a given fraction of the incident spectrum into a power law shape with a photon index $\Gamma$. The amount of the up-scattered fraction of the incident radiation is stored in a parameter called scattered fraction. We allowed for a reflection component and modeled it with \texttt{reflect} \citep{1995MNRAS.273..837M}. The reflection component affected solely the up-scattered photons \citep[see also][]{2011ApJ...742...85G,2012ApJ...753...65T}. As \texttt{simpl} and \texttt{reflect} are convolution models it is necessary to calculate the model well-outside the normal bounds of the PCA and HEXTE energy range. Therefore, we used an extended energy range from 0.1 keV to 1 MeV for our fits. The distribution of reflection values for all three sources are shown in Fig.\ \ref{Fig:refl}. The mean error is about 0.15 (\gx339, \xte) to 0.2 (\h1743), although individual observations can show much smaller or bigger errors. If needed a high-energy cut-off (\texttt{highecut}) was included. This was mostly the case for observations observed during the rise of an outburst. We added a Gaussian to account for excess emission at 6.4 keV. The centroid was allowed to vary between 6.4 and 6.7 keV and the line width was constrained between 0 and 1 keV to prevent artificial broadening due to the response of the PCA detector at 6.4 keV. For the foreground absorption we used the \texttt{TBabs} model \citep{2000ApJ...542..914W}, with fixed foreground absorption ($N_{\mr{H}}$). Values used for the individual sources are: 5.0\hcm{21} for \gx339 \citep{2004MNRAS.351..791Z}, 1.6\hcm{22} for \h1743 \citep{2009MNRAS.398.1194C}, and 6.7\hcm{21} for \xte\ \citep{2004ApJ...601..439T}.
For most observations the disc blackbody temperature lies between 0.5 and 1.2 keV and we obtain an inner disc radius of a few tens kilometers. The reduced $\chi^2$ value, $\chi^2_{\mr{red}} = \chi^2/N_{\mr{dof}}$, where $N_{\mr{dof}}$ is the number of degrees of freedom (dof), is less than or around one for most of the observations. For a small fraction ($\approx2$\%) of spectra the value of  $\chi^2_{\mr{red}}$ exceeds 1.5.\@ However, it never reaches a rejection limit of two.

\begin{figure*}
\resizebox{\hsize}{!}{\includegraphics[clip,angle=0]{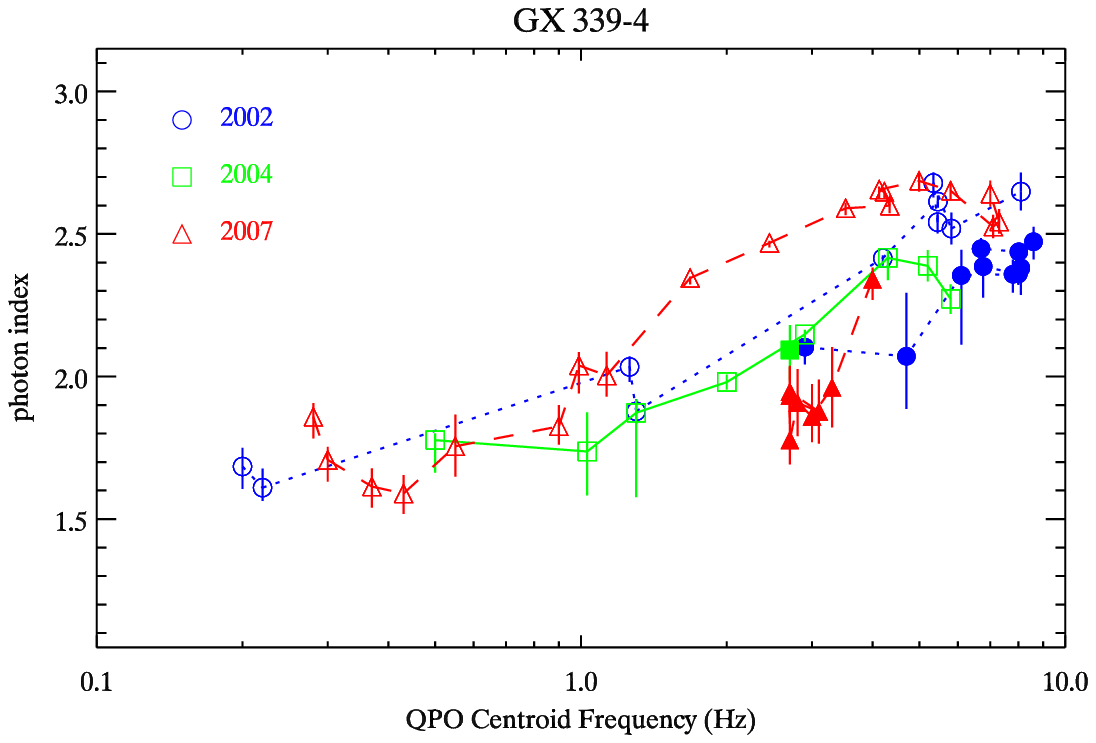}\hskip0.2cm\includegraphics[clip]{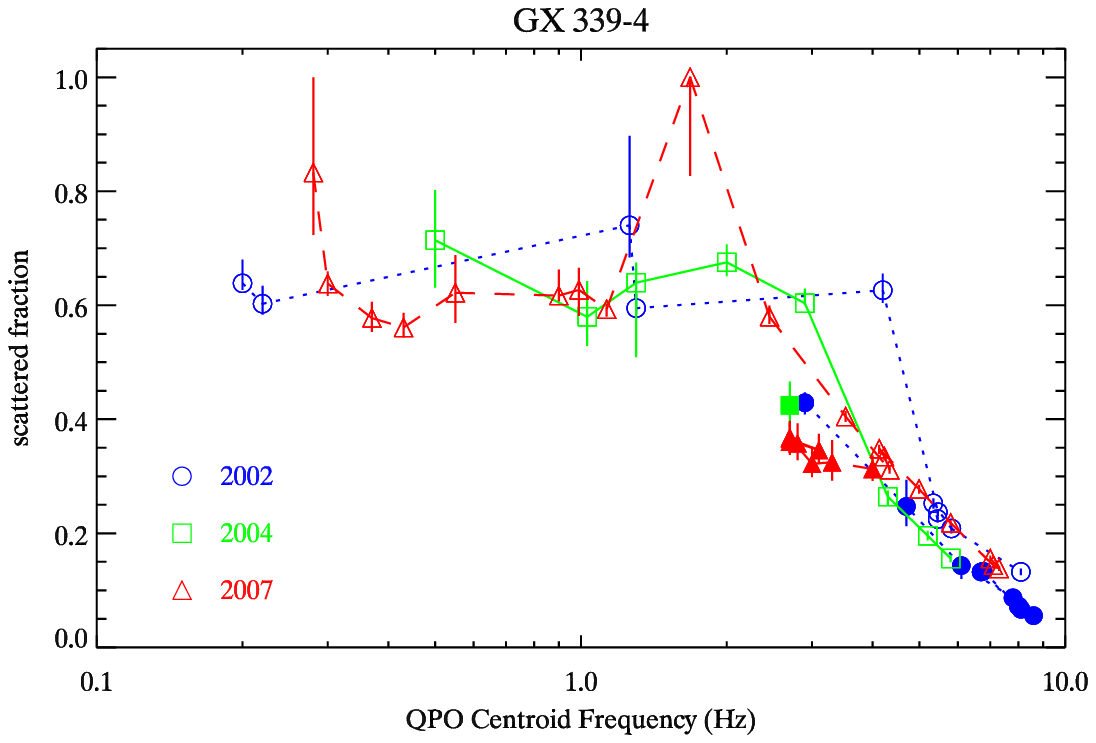}}
\resizebox{\hsize}{!}{\includegraphics[clip,angle=0]{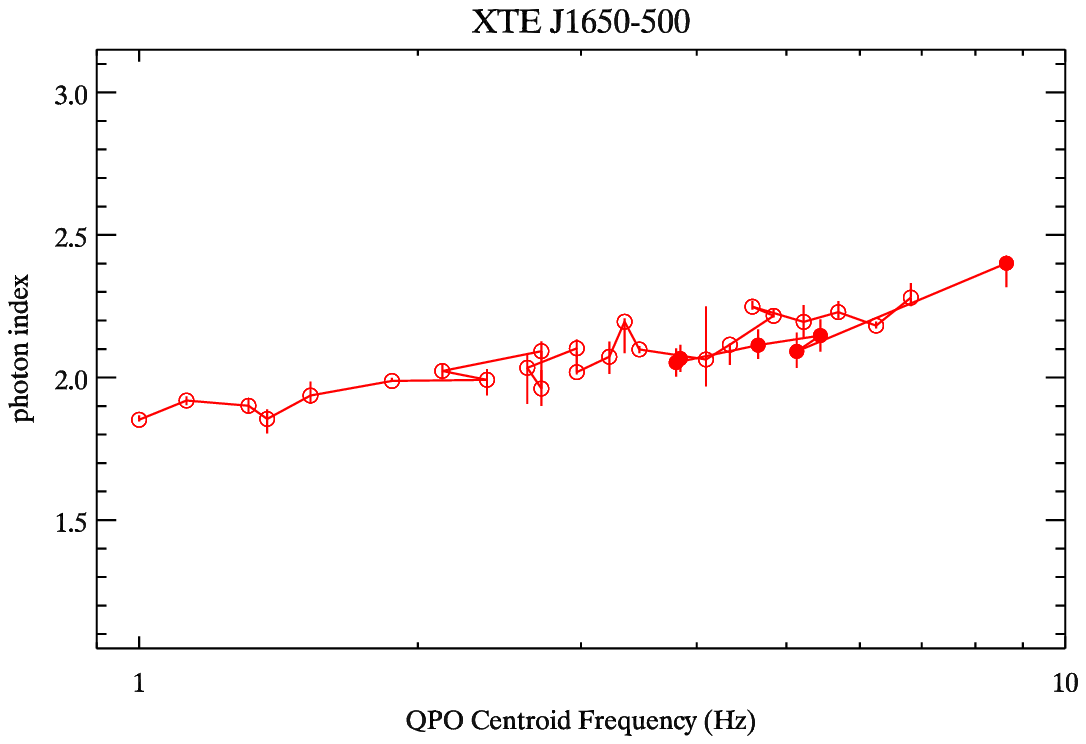}\hskip0.2cm\includegraphics[clip]{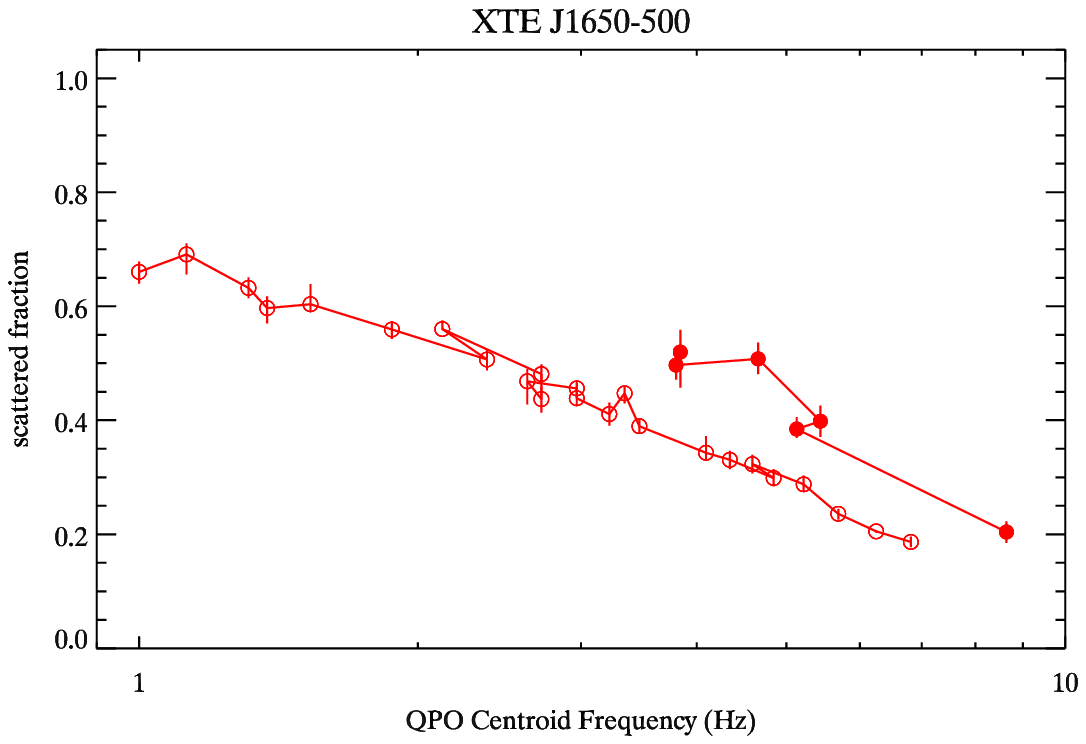}}
\resizebox{\hsize}{!}{\includegraphics[clip,angle=0]{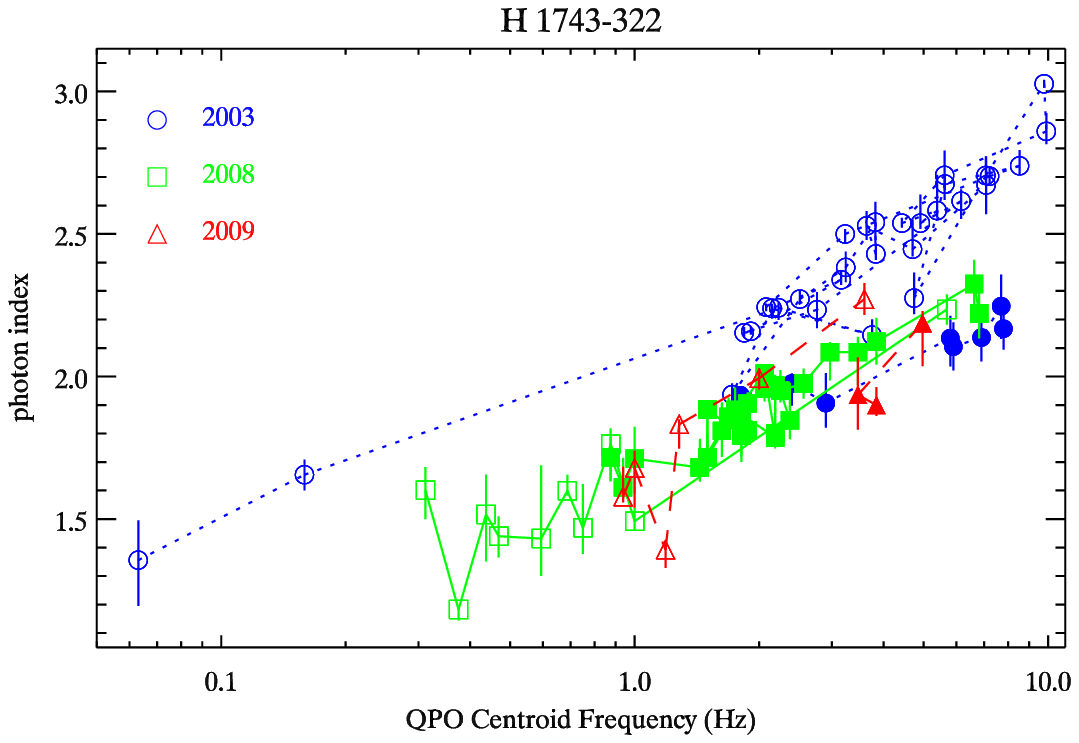}\hskip0.2cm\includegraphics[clip]{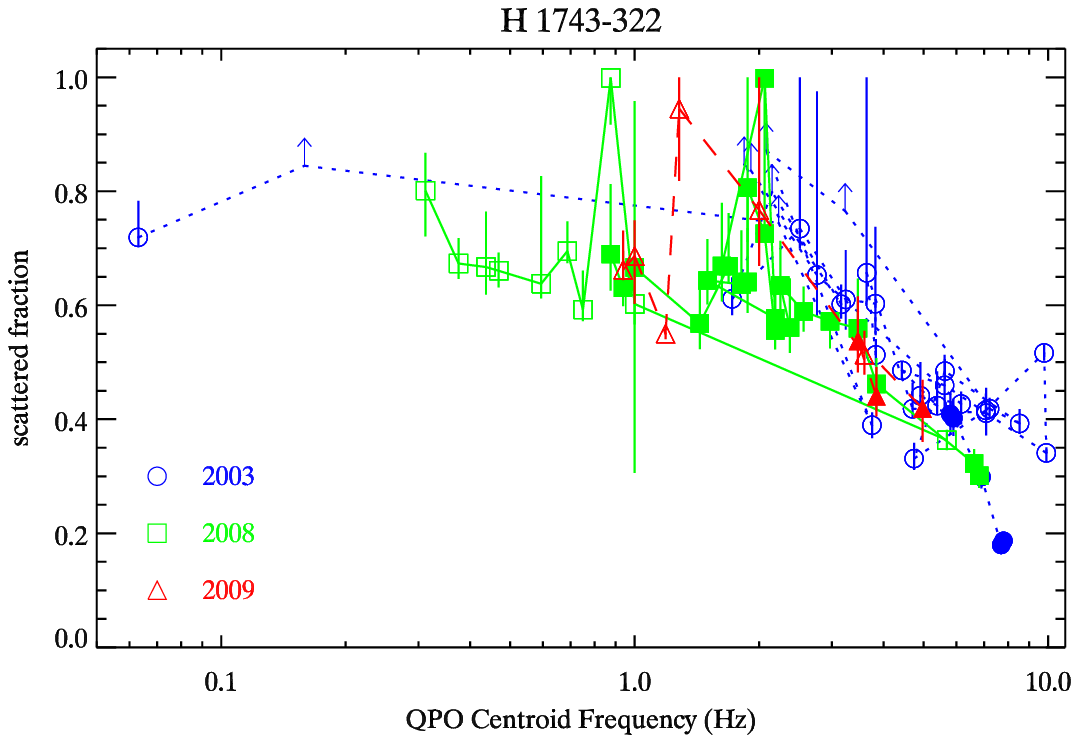}}
\caption{$\Gamma$-QPO (left column) and SF-QPO (right column) correlations for \gx339\ (upper row), \xte\ (middle row), and \h1743\ (lower row). Open symbols denote observations belonging to the rise branch of an outburst, while filled symbols mark observations of the decay branch. If a source has been detected during more than one outburst, the individual outbursts are marked with different symbols. Observations for which we obtained only a lower limit of the scattered fraction are marked by arrows.}
\label{Fig:corr}
\end{figure*}

\begin{figure*}
\resizebox{\hsize}{!}{\includegraphics[clip,angle=0]{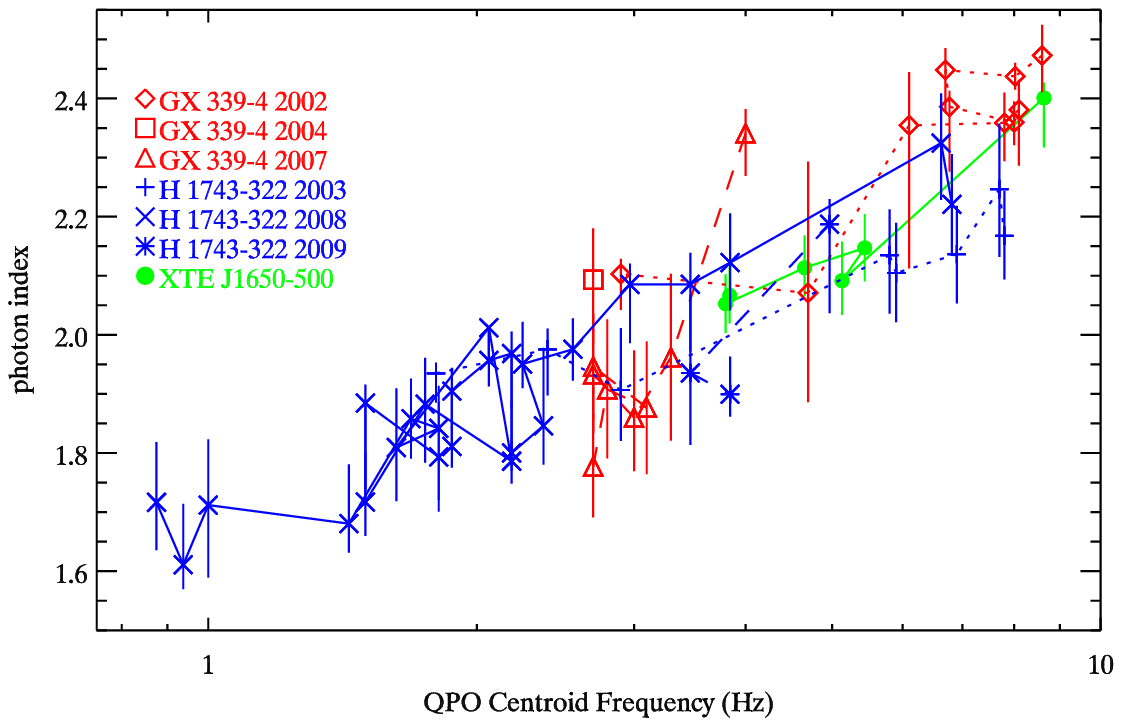}\hskip0.2cm\includegraphics[clip]{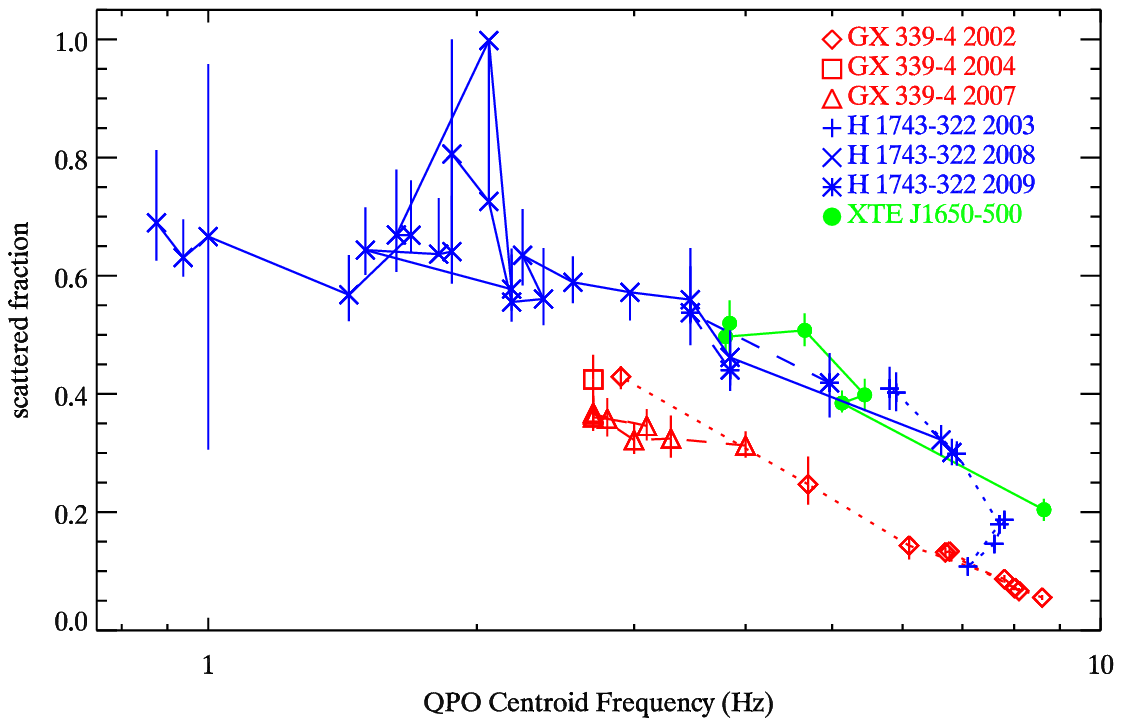}}
\caption{The $\Gamma$-QPO (left column) and SF-QPO (right column) relation during decay for all seven outbursts. In the $\Gamma$-QPO diagram all points seem to follow the same track, without recognisable differences between individual sources. In the SF-QPO diagram there are two different tracks. The points obtained from observations of \h1743\ and \xte\ are located on one track, while those of \gx339\ span their own track at lower scattered fractions.}
\label{Fig:decay}
\end{figure*}

\section[]{Correlations}
\label{Sec:res}
In this section we present correlations between QPO centroid frequency and different spectral parameters. Figure~\ref{Fig:corr} shows for the three BHTs the correlations between QPO centroid frequency and photon index or scattered fraction, respectively. Individual outbursts of the same source are marked with different symbols. The broad distribution of reflection fractions (see Fig.\ \ref{Fig:refl}) and the rather large systematic errors hamper the detection of any clear trends in the correlation between reflection fraction and QPO frequency. We do not show correlations between QPO frequencies and disc parameters, as the obtained disc parameters should be taken with care. The working range of PCA (3 -- 40 keV) covers only the high energy part of the disc component, above the Wien peak. The missing coverage of lower energies together with the presence of a strong Comptonized component increases the uncertainties in the derived disc parameters, especially for observations obtained at the very begin and end of an outburst. In addition, it is known that the spectral parameters derived from the \texttt{diskbb} model should not be interpreted literally \citep[see \eg\ ][]{2000MNRAS.313..193M,2006ARA&A..44...49R}. Nevertheless, we want to mention that our investigations imply that QPOs at the same frequency occur at a lower inner disc radius in the decay branch compared to the rise branch.

\subsection{The scattered fraction -- QPO centroid frequency relation (SF-QPO relation)}
All investigated outbursts show a negative correlation between the fraction of up-scattered photons and the QPO centroid frequency: a high scattered fraction is observed at low QPO frequencies, and the scattered fraction decreases with increasing frequency. In the case of \xte\ the correlation is most obvious and we recognise two branches which belong to the rise and decay branch of the outburst, where the latter one lies above the branch of the rise. For the other two sources the distinction between rise and decay is less clear. 
For \h1743\ it is impossible to separate different branches. For a few observations we obtained a scattered fraction close to unity. In these cases we used the 1$\sigma$ lower error as a lower limit (indicated in Fig.\,\ref{Fig:corr} by a small arrow). 

\subsection{The photon index -- QPO centroid frequency relation ($\Gamma$-QPO relation)}
The photon index rises with increasing QPO centroid frequency. In the case of \xte\ it is a shallow and rather linear rise. The two branches belonging to the rise and decay of the outburst are indistinguishable. We recognise distinct rise and decay branches for the 2002 and 2007 outburst of \gx339. The branch belonging to the rise of the 2004 outburst seems to be aligned with the decay branch of the 2002 outburst. This might be related to the lower luminosity at which the 2004 outburst has been observed. The photon indices are well constrained. The $\Gamma$-QPO relation in \gx339\ can be described as a loop-like structure compared to the narrower, more band-like appearance in the case of \h1743. 

\subsection{$\Gamma$-QPO relation during outburst decay}
In Fig.\,\ref{Fig:decay} (left column) we summarised in one diagram the $\Gamma$-QPO relation during decay for all outbursts. This figure reveals that the $\Gamma$-QPO relation follows a general track during outburst decay. The relation can be approximated by $\Gamma=\Gamma_0+a*\mr{cf}_{\mr{QPO}}$, with $\Gamma_0=1.73$ and $a=0.08$. The standard deviation in $\Gamma$ is $0.10$. We would like to point out that during outburst decay also the distribution of source luminosities is narrower. 

\subsection{SF-QPO relation during outburst decay}
In Fig.\,\ref{Fig:decay} (right column) we summarised in one diagram the SF-QPO relation during decay for all outbursts. Unlike the $\Gamma$-QPO relation, the SF-QPO relation shows two tracks. The upper one consists of all observations of \xte\ and of most observations of \h1743. The lower track contains the observations of \gx339.  Only, the 2003 outburst of \h1743\ starts at $\sim$7 Hz on the correlation of \gx339, leaves it and reaches the upper track spanned by the other observations of \h1743.

\section[]{Discussion}
\label{Sec:dis}
We studied for the BHTs \gx339, \h1743, and \xte\ the correlation between the QPO centroid frequency and two spectral parameters: the photon index and the fraction of up-scattered photons. The correlations presented in this work agree qualitatively with the ones that can be derived from the values given in Table~2 of \citet{2009ApJ...699..453S}. This means that the anti-correlation in the SF-QPO relation and the positive correlation in the $\Gamma$-QPO relation are model independent. We note that the photon indices in the present work are slightly higher than those derived with the bmc model. This has to be expected, as we included a reflection component in our spectral model. \citet{2003A&A...397..729V} investigated the correlation between QPO frequency and photon index for an additional five sources, partially taking QPO frequencies and spectral parameters from the literature. A single correlation between spectral index and QPO centroid frequencies was also found by \citet{2002PhDT........12K} during the outburst decays of six sources in eight outbursts.  

X-ray spectra of black hole X-ray binaries consist of two main components: The soft component is believed to originate in the geometrically thin and optically thick Shakura-Sunyaev accretion disc \citep{1973A&A....24..337S}.  One of the most plausible processes of formation of the hard spectral component is Comptonization of soft disc photons on hot electrons \citep{1979Natur.279..506S,1980A&A....86..121S}. The Comptonization site is often referred to as a corona. Although it is  generally accepted that the Comptonizing corona has to be located in the close vicinity of the black hole, there are currently different ideas brought up on the detailed geometry of the region. The overall behaviour of the SF-QPO and $\Gamma$-QPO relations can be explained qualitatively within the ``sombrero'' geometry \citep{1997MNRAS.292L..21P,2010LNP...794...17G}. In this configuration, a quasi-spherical corona surrounds the black hole and the accretion disc extends a short distance into the corona. In the LHS the disc is truncated at a large radius and the remaining space is filled by the hot, optically thin corona \citep{2007A&ARv..15....1D,2008MNRAS.388..753G,2008A&A...488..441D,2009ApJ...707L..87T,2009MNRAS.394.2080H}. While during the HSS, the disc reaches the inner most stable orbit leaving (at most) only little space for the corona. A fraction of the Comptonized photons emitted from the corona irradiates the accretion disc. Some of them are reflected due to Compton Scattering \citep{1974A&A....31..249B}. We note that there are claims that the disc reaches the inner most stable orbit even in the LHS \citep{2006ApJ...652L.113M,2006ApJ...653..525M,2007ApJ...666.1129R}. There are also models which predict a hot inner and a cool outer disc separated by a gap filled with an advection-dominated accretion flow \citep{2007ApJ...671..695L}.

Coming back to the sombrero geometry, which matches our findings, the QPOs correspond to oscillations in a transition layer between the disc and the hotter Comptonizing region \citep{2000ApJ...538L.137N,2009MNRAS.397L.101I,2011MNRAS.415.2323I}.  The QPO frequency is inversely related to the truncation radius of the disc, as it is in most models \citep[see \eg][]{2000ApJ...542L.111T,2000ApJ...531L..41C}. While the system evolves from the LHS to the HIMS the truncation radius moves inward and the QPO frequency increases. In the sombrero configuration the solid angle of the disc seen by the hot electrons correlate with the fraction of disc emission reaching the corona. As the disc expands towards the black hole, the reflection scaling factor increases, the fraction of up-scattered photons decreases and the spectrum steepens \citep{1999MNRAS.303L..11Z,2010LNP...794...17G}.  Thus the observed (anti-)correlations between QPO frequency and spectral parameters can be explained within the sombrero geometry. 
 
\section{Conclusion}
We studied correlations between spectral and timing parameters for observations with type-C QPOs. The sample comprised observations of \gx339, \h1743, and \xte. Our investigations confirmed the known positive correlation between photon index and centroid frequency of the QPOs and revealed an anti-correlation between the fraction of up-scattered photons and the QPO frequency. We showed that both correlations behaved as expected in the ``sombrero" geometry, which also predicted the observed correlation between photon index and reflection scaling factor \citep{2010LNP...794...17G}. Furthermore, we showed that during outburst decay the correlation between photon index and QPO frequency followed a general track, independent of individual outbursts. 

\section*{Acknowledgments}
The research leading to these results has received funding from the European Community's Seventh Framework Programme (FP7/2007-2013) under grant agreement number ITN 215212 ``Black Hole Universe". SM and TB acknowledge support from grant ASI-INAF I/009/10/0.
This work makes use of EURO-VO software, tools or services. The EURO-VO has been funded by the European Commission through contracts RI031675 (DCA) and 011892 (VO-TECH) under the 6th Framework Programme and contracts 212104 (AIDA) and 261541 (VO-ICE) under the 7th Framework Programme.

\bibliographystyle{mn2e}
\bibliography{/Users/apple/work/papers/my2010}

%\appendix

\bsp

\label{lastpage}

\end{document}